# Theoretical insights on the importance of anchoring vs molecular geometry in magnetic molecules acting as junctions

Silvia Giménez-Santamarina[1], Salvador Cardona-Serra[1], Alejandro Gaita-Ariño[1]

[1]ICMol. Universitat de València. C/ Catedrático José Beltrán nº 2, 46980 Paterna, Valencia, España.

**Abstract**

The anchoring of the molecule to an electrode is known to be a key factor in single-molecule spintronics experiments. Likewise, a relaxation down to the most stable geometry is a critical step in theoretical simulations of transport through single-molecule junctions. Herein we present a set of calculations designed to analyze and compare the effect of different anchoring points and the effect of perturbations in the molecular geometry and interelectrode distance. As model system we chose the $[V(\alpha-C_3S_5)_3]^{2-}$ complex connecting two Au(111) electrodes in a slightly compressed geometry. In our calculations, the attachment happens through an S-Au bond, a common anchoring strategy in molecular spintronics experiments. Our results confirm that small alterations in the molecular geometry have important effects in the conductance. We were able to compare these effects with the ones arising from changing the anchoring position with a constant molecular geometry. Unexpectedly, we demonstrate that the anchoring position has only a lesser relevance in the spintronic behavior of the device, as long as all other parameters are kept frozen. As a consequence, we predict that for experimentalists aiming for reproducibility, the molecular design of rigid linkers is more relevant than the design of univocal anchoring positions.

Introduction

Spintronics has been consolidated as a scientific field capable of originating major technological advances since the discoveries of giant magnetoresistance[1] and tunnel magnetoresistance[2]. In spintronics, the use of magnetic materials in addition to electric conductors allows information to be encoded not only as charge flow but also as spin flow[3]. Recent milestones include the use of conventional inorganic magnetic tunnel junctions to build nanoscale nonlinear oscillators[4] and memristive ferroelectric nanomaterials[5], both of which bring closer the development of high-density, low-power neuromorphic-based computing[6,7,8].

Within Spintronics, the development of Organic Spintronics means addressing the challenge of designing and obtaining molecule-based materials that imitate or improve the behavior of traditional inorganic spintronic materials. These materials are intended to present specific magnetic and electric behavior as well as chemical stability to guarantee reproducibility, and consequently, allow its processing as spintronic device. The miniaturization limit of molecular spintronics is single-molecule spintronics, where the goal is to obtain molecules that function as 'smart' building blocks to construct spintronic devices from the bottom up. While single-molecule experiments are relatively scarce so far in Spintronics, these experiments already have resulted in unique opportunities for the exploration of quantum behavior. In a germinal example[9], the transport properties of a bis-pthalocyaninate-$Tb^{3+}$ complex in a break junction setup allowed to probe the quantum state of a single nuclear spin. An extension of this experiment resulted in a 2-qubit version of Grover's quantum algorithm[10]; analogous 3-qubit systems have already been proposed[11,12,13].

Currently, the main limitation of experiments in single-molecule spintronics is their poor reproducibility. In turn, this is mainly attributed to the non-reproducible attachment between the molecules and the electrodes. It is known that small differences in the attachment or in the electrode surface can have critical consequences in the electric and magnetic response of a molecular-based device[14]. This is important since most single molecule transport experiments are inherently prone to at least minor irreproducibilities in the microscopic structure of the electrode surfaces and in the molecular anchoring.[14] Indeed, it is widely accepted that to better focus synthetic and experimental efforts we need a better understanding of the effect of the structural adsorption geometry in single molecule transport experiments[15]. A wide range of theoretical studies has been performed in the last two decades to analyze this effect. Most of them use the rigid benzene-dithiol (BDT) as a model molecule and NEGF-DFT calculations,[15, 16, 17] with the recent addition of atomic self-interacting correction (ASIC) for the proper alignment of the molecular orbitals with the Fermi energy of the extended electrode orbitals[18].

In the present work we aim to estimate the effect of the anchoring position, which is often assumed to be key to the transmission spectra, and estimate their relative importance compared with the effects of structural distortions and interelectrode distance. We employ NEGF-DFT-ASIC and chose the tris-dithiolate complex $[V(\alpha-C_3S_5)_3]^{2-}$ sandwiched between two Au (111) surfaces as a model system.

Theoretical Methodology

First-principle calculations are performed using the SMEAGOL code[19] that interfaces the non-equilibrium Green's function (NEGF) approach to electron transport with the density functional theory (DFT) package SIESTA[20] (see more details on SI). In our simulations the transport junction is a two-terminal molecular device consisting of a complex $[V(\alpha-C_3S_5)_3]^{2-}$ between two Au(111)-oriented surfaces acting as electrodes, mimicking a standard transport break-junction experiment with the most used gold surface orientation. As model system we chose the tris-dithiolate. We previously explored this vanadium(IV) tris-dithiolate complex in the context of nuclear-spin-sensitive single-molecule spin transistors. In the present work we employed 7x7 surfaces rather than 5x5 as in previous works, since we wanted to allow a single-molecule relaxation with no intermolecular packing effects.

We employed different starting anchoring positions to explore a reasonable variation of the estimated transport properties in imperfect experimental conditions (see Figures 1, 2 and Supporting Information for details). For these purposes, we considered cases with a terminal sulfur at the "adatom" position (on top of a gold atom), or on top of an edge, or in proximity to the so-called HCP (FCC) positions, which refer to the triangle centers which have (not) a gold atom in the layer directly beneath them. We label our models as "adatom" (**A**), "on-edge" (**B**), "off-centered triangle I" (**C**) and "off-centered triangle II" (**D**). The initial distance between the sulfur nucleus and the gold plane was set to 2.0 Å as a reasonable assumption according to the literature[21]. The near $C_3$ geometry was maintained for each model (see details in Supporting Information). Preserving control over the starting molecular geometry allows us to independently evaluate the effect of geometrical distortions and anchoring positions. In the calculations we apply geometry constrictions to electrodes with the objective of simulating fixed electrodes. Interelectrode distance is kept for each of these four models at 17.1 Å.

Two extra models were included in the study giving an extra degree of freedom. In the first extra model, non-symmetric distances S-Au$_{plane}$ were chosen (2 Å and 2.5 Å), with an "off-centered triangle I" anchoring position (**E**): the interelectrode distance was thus increased to 17.6 Å. For the second extra model, we chose the "adatom" anchoring (**F**) but applied a 120° rotation to the molecule. Because of the slight deviation from $C_3$ symmetry, applying this rotation while keeping a S-Au$_{plane}$ of 2 Å distance results in a final interelectrode distance of 16.7 Å (See Structural details in Table S1). We aim with those extra models to evaluate transport spectra while the molecule is subjected to different degrees of structural stress.

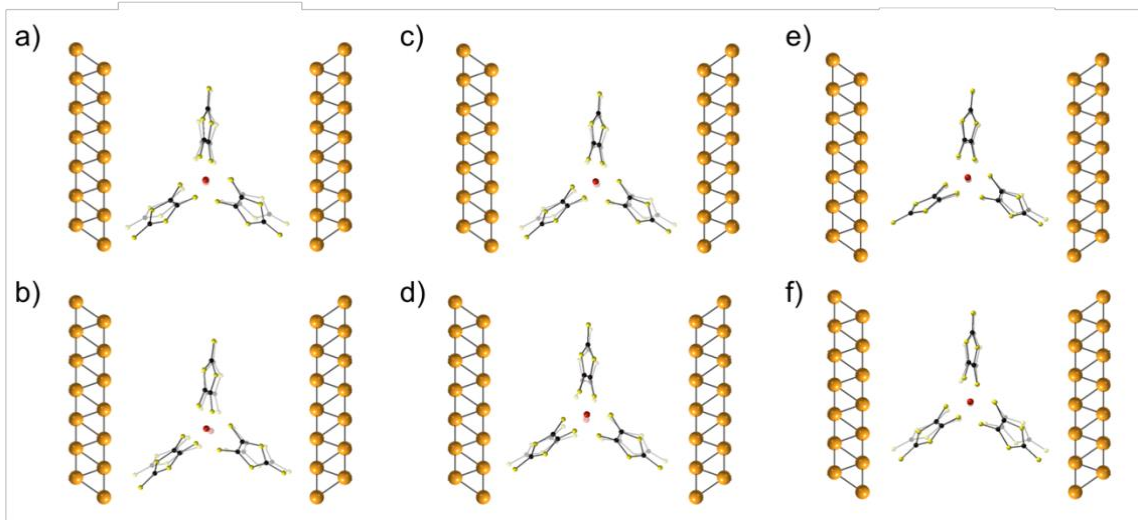

Figure 1: Scheme of the two terminal molecular junction models. Visual description of the tree-dimensional structure before (transparent) and after (opaque) relaxation of **A** ("adatom"), **B** ("on-edge"), **C** ("off-centered I"), **D** ("off-centered II"), **E** ("non-symmetrical off-centered I") and **F** ("rotated adatom").

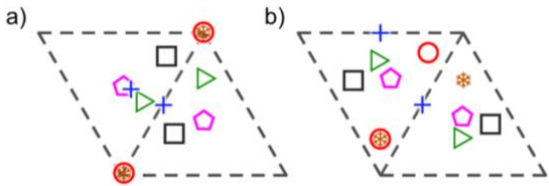

Figure 2: Schematic representation of the (a) pre-relaxation and (b) post-relaxation attachment positions of the two sulfur atoms: **A** ("adatom", red circles), **B** ("on-edge", blue crosses), **C** ("off-centered I", green small triangles), **D** ("off-centered II", black squares), **E** ("non-symmetrical off-centered I", pentagons) and **F** ("rotated adatom", snowflakes). The vertices of the triangles represent gold atoms. The triangle pointing down represents an HCP hollow site, with the triangle pointing up representing an FCC hollow site.

Results

Let us start by analyzing the transmission spectra prior to relaxation (see Figure 3a). The six transmission spectra share the same features, in terms of number of transmission peaks, position and intensities. In all cases the spectra present a double peak from majority carriers at the Fermi energy ($E_F$) with almost no contribution of minority carriers. Coincident with a prior work the molecules are expected to present good conductance at small (0.5 V) bias voltages, as well as a strong spin-up filtering,[11] if the spin in the vanadium is polarized by an external magnetic field among the four cases with identical molecular structure and interelectrode distance. The case with anchoring on the "adatom" (**A**) shows the greatest discrepancies, which amount to deviations of about 0.1 eV in the position of some transmission peaks, whereas the similarity is maximal between the two "off-centered triangles" models (**C** and **D**). This correlates well with similarities and differences in anchoring geometries. In terms of energy, the only clear correlation is in the case of the "adatom" model (**A**), which stands out as more unstable than the other three by about 6 eV. Cases **E** ("non-symmetrical off-centered I") and **F** ("rotated adatom") display transmission spectra that, while recognizable, are better differentiated from the first four cases. This signals the influence of the interelectrode distance (and, in case **F**, of the molecular distortions) in transport properties, over the influence of the anchoring point. The calculated energies evidence the importance of the sulfur-gold compression and are not correlated with transport properties. Case **E**, where one of the distances is 2.5 Å rather than 2.0 Å, has the lowest pre-relaxation energy, and case **F** has a similar pre-relaxation energy as case **A**, since they are both "adatom" cases with a direct sulfur-gold contact that is much too close.

The spectra of the six relaxed structures (see Figure 3b) are relatively similar to each other and distinct from the pre-relaxation spectra. In all cases, features are distorted but recognizable. All spin-up and spin-down peaks survive the relaxation process, although with shifts of about 0.3 eV and some important changes in peak intensity. In all cases the molecule would still be expected to present conductance at small (0.5 V) bias voltages, as well as a strong spin-up filtering, if the spin in the vanadium is polarized by an external magnetic field. The results of cases **E** and **F** are revealing. Case **E**, being unique in its very different initial sulfur-gold distances, relaxes into a more distinct molecular geometry which results in the most differentiated transmission spectrum. Case **F**, having started in the same anchoring point as case **A**, relaxes to a very similar geometry. As a result, and despite presenting different interelectrode distances, post-relaxation transmission spectra of cases **A** and **F** are nearly identical. This again highlights the preponderance of the molecular geometry in the transmission spectrum, over the attachment and the interelectrode distance which only have an indirect effect.

For completeness, we add the comparison with a transmission spectrum presented by some of us in a previous publication for the same molecular system.[11] The main structural difference is that in the previous case was relaxed between 5x5 Au atoms while in the present case we are using 7x7. With this, the interaction with other molecules due to periodic boundary conditions is lowered in the present calculation. This effect has a small influence in the transport properties: the main two spin up peaks at EF were also present, although almost completely overlapped. The rest of the spectra is almost maintained except for some spin-up peaks at negative energies which where sharp and now are almost inexistent.

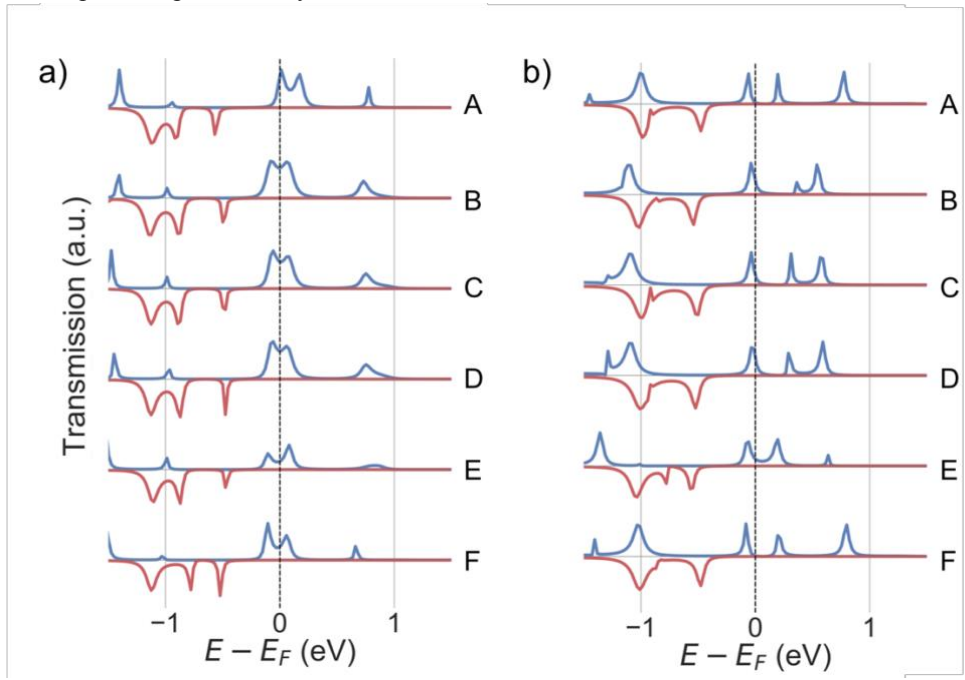

Figure 3: From top to bottom: Transmission spectra of **A** ("adatom"), **B** ("on-edge"), **C** ("off-centered I"), **D** ("off-centered II"), **E** ("non-symmetrical off-centered I") and **F** ("rotated adatom") respectively. For the spectra on the left (a), the molecular structures were frozen. For the spectra on the right (b), each molecular structure had been individually relaxed.

The relaxation starting from different situations allowed a complementary study of preferential anchoring positions. We performed the geometrical relaxation of the six proposed structures by permitting the molecule to relax over a fixed Au(111) perfect layer.

All relaxations involved relatively small displacements (<1.5 Å), and the variety and low-symmetry nature of the final anchoring positions (see Figure 2) do not evidence any strong preference for a perfect adsorption point neither a clear preference between HCP hollow site (gold atom directly beneath its center) or FCC hollow site (no gold atom directly beneath its center). This is probably related with the fact that the complex needs to accommodate a rigid interelectrode distance by means of molecular distortions. This means that in the equilibrium geometry any further stabilization of one of the anchoring points means destabilizing the other and/or straining the molecular structure.

Distances between the molecule and the substrate (from sulfur nucleus of each endpoint and the gold plane) after relaxation shows an average value of 2.56 Å, with a window of minimum and maximum values from 2.48 to 2.62 Å respectively. Note that the initial distances were 2 Å except in one of the sides of non-symmetric case E. Thus, there is a certain compression in all cases which had to be alleviated by a distortion of the complex.

The relative relaxation energies for each structure are shown in Figure 4. Comparing the first four models where only the anchoring position was changed, the most stable case was identified as the S atom located at the "on-edge" (**B**) site, the relaxation energy was $\Delta_B = 1.44$ eV. A slightly higher change is found when the attachment point starts in the "off-centered triangle I" (**C**) ($\Delta_C = 1.76$ eV). Furthermore, when the S atom (due to relative proximity) tends to be pinned to the meta-stable "off-centered triangle II" (**D**), the relaxation energy is $\Delta_D = 2.41$ eV. A most extreme case is obtained when the S is located directly over a gold atom ('adatom', **A**) where the relaxation involves a huge energetic stabilization ($\Delta_A = 6.06$ eV).

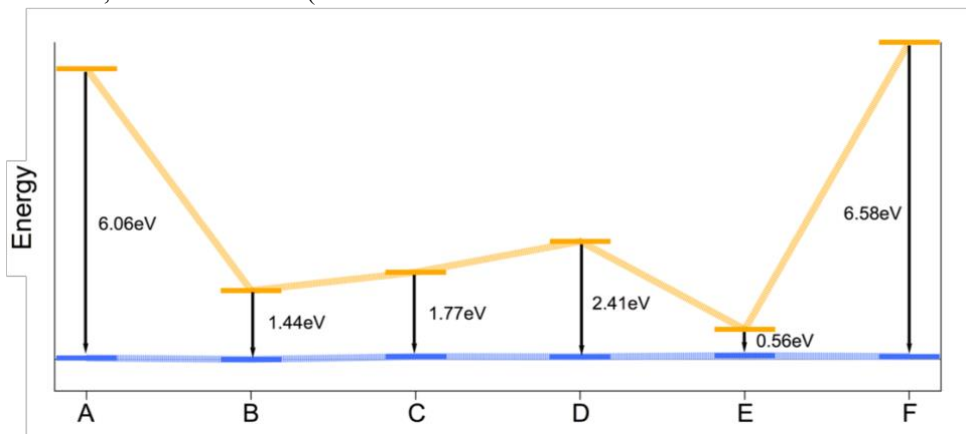

Figure 4: Relative energies before (yellow lines) and after (blue lines) the relaxation process. Energies after relaxation are not normalized to zero, but are very similar to each other after relieving the tension of the close sulfur-gold contacts.

Analyzing the relaxed configuration for all the systems that were originally isostructural, all energies are within a window typical for different chemical conformations: 0.030 eV, 0.060 eV and 0.056 eV for models **A**, **C** and **D** respectively compared with the most-stable case, **B**. Model **E**, with less strain, is the most stable configuration before geometrical optimization, whereas model **F**, with more strain and an "adatom" anchoring position ($\Delta_F = 6.58$ eV); after optimization, **F** has a lower energy than **E**, with $\Delta_{EF} = 0.016$ eV.

Concluding remarks

The details governing single-molecule spintronics experiments are not yet fully understood. Many magnetic molecules have promising properties in bulk but their usefulness in the device may depend on factors such as their rigidity or in the nature of their anchoring groups. In this work we presented a set of calculations designed to analyze the influence in electronic transport of different anchoring points and to compare it with the effect of perturbations in the molecular geometry and interelectrode distance. We have performed DFT-NEGF calculations on a $[V(\alpha-C_3S_5)_3]^{2-}$ complex connecting two Au(111) electrodes to obtain the transmission spectra of a total of 12 structures, differing in the anchoring point, the molecular geometry and/or the interelectrode distance. Contrary to the consensus, our results indicate that the anchoring between molecule and electrode has a relatively small influence in single-molecule spintronics experiments; of course, for structural optimizations it is important to start from a more stable anchoring position to save computing time. In contrast, our results confirm that relatively small distortions in the molecular geometry may dominate the conductance. The importance of molecular flexibility could not be detected in previous systematic studies which employed benzene-1,4-dithiol as model system, but it can be crucial for experiments with magnetic complexes, which typically have structural flexibility.[9] This means that a reproducible anchoring to the electrode plus a flexible bridge that guarantees the pristine single-molecule properties might not be ideal, since the uncontrolled bending of this bridge could actually govern the transport properties. For the molecular design and for device preparation, our results suggest that achieving a unique anchoring point is not crucial and that the focus should be put in a very robust molecular geometry.

Acknowledgments

The present work has been funded by the EU (ERC Consolidator Grant 647301 DECRESIM, and COST 15128 Molecular Spintronics Project), the Spanish MINECO (grants MAT2017-89528 and CTQ2017-89993 cofinanced by FEDER and Excellence Unit

María de Maeztu MDM-2015-0538), and the Generalitat Valenciana (Prometeo Program of Excellence). A.G.-A. acknowledges funding by the MINECO (Ramón y Cajal Program) S.C.-S. thanks the Spanish MINECO for a Juan de la Cierva Incorporación postdoctoral Fellowship.

# Supporting information for:

# Theoretical insights on the importance of anchoring vs molecular geometry in magnetic molecules acting as junctions.

Silvia Giménez-Santamarina[1], Salvador Cardona-Serra[1], Alejandro Gaita-Ariño[1]

[1]*ICMol. Universitat de València. C/ Catedrático José Beltrán nº 2, 46980 Paterna, Valencia, España.*

Contents:

1. Details on the structural models.
   a. Table 1.- Summary of geometrical parameters describing each model after and before the structural relaxation. The angles $\Theta_1$, $\Theta_2$ and $\Theta_3$ are illustrated in Figure S1.
2. Theoretical methodology.

1. Details on the structural models.

    a. Initial models

We proposed different attachment points between the complex [V(α-C$_3$S$_5$)$_3$]$^{2-}$ and two Au(111)-oriented surfaces acting as electrodes. The structural coordinates of the molecule are obtained from a previous calculation, see Cardona-Serra et al. J. Phys. Chem. Lett. 2017. We have considered different cases in which the terminal sulfur atoms are:

Model **A**: the "*adatom*" position (on top of a gold atom). One sulfur is almost perfectly on top of a gold atom and the other is slightly deviated towards an FCC position.

Model **B**: "*on-edge*", on top of an edge. One sulfur is on almost perfectly on top whereas the other is deviated towards an HCP position.

Model **C**: "*off-centered triangle I*". One sulfur is between an edge and an FCC position, the other is almost on an edge but slightly deviated towards an HCP position.

Model **D**: "*off-centered triangle II*". One sulfur is between an edge and an FCC position, the other is between an edge, an HCP position and an "*adatom*" position.

Model **E**: with non-symmetric distance between terminal sulfur and gold plane. Non-symmetric distances S-Au$_{plane}$ were set at 2 Å and 2.5 Å. The anchoring position was "*off-centered triangle I*".

Model **F**: where the molecule was rotated 120° along the C$_3$ axis. The anchoring position was "*adatom*".

The initial distance between the sulfur nucleus and the gold plane for models A, B, C and D was set to 2.0 Å as a reasonable assumption according to the literature[1] and interelectrode distance was kept at 17.12 Å. In model E, the distance between one of the terminal sulfur atoms to the nearest gold plane was increased by 0.5 Å, but geometrical structure was maintained. For model F, the interplanar gold distance is lowered by 0.5 Å as a consequence of the 120° rotation.

    b. Obtained relaxed models

After geometrical relaxation:

Model **A**, "*adatom*": evidenced the instability of the adatom anchoring points. After relaxation both anchoring points are between an edge, an HCP position and an adatom position.

In model **B**, "*on-edge*": after relaxation, the first sulfur atom is still on the center of an edge whereas the second is now closer to the center of an edge but slightly deviated towards an "*adatom*" position.

In model **C**, "*off-centered triangles I*": after relaxation, the first sulfur is now almost perfectly on an FCC position, whereas the other is between an edge and an HCP position.

In model **D**, "*off-centered triangles II*": after relaxation, the first sulfur is now less deviated from being on top of an edge, whereas the second is less deviated from being on top of an HCP position.

In model **E**, non-symmetric: after relaxation, both sulfurs are now less deviated from being on the center of the triangles.

In model **F**, "*adatom*" rotated molecule: after relaxation, both sulfurs are slightly attracted to the center of the triangles, and equivalently located in the middle point of the angle of the vertex of the starting "*adatom*" position.

Distances between the molecule and the substrate for all relaxed structures show an increase with an average value of 2.56 Å. A summary of these results can be found in Table S1.

**Table S1:** Summary of geometrical parameters describing each model after and before the structural relaxation. The angles $\Theta_1$, $\Theta_2$ and $\Theta_3$ are illustrated in Figure S1.

| Code | Initial $D_{S-Au}$ (Å) | Final $D_{S-Au}$ (Å) | Angle $\Theta_1$ (degrees) | Angle $\Theta_2$ (degrees) | Angle $\Theta_3$ (degrees) | $D_{interplanar}$ (Å) | $E_{initial}$ (eV) | $E_{final}$ (eV) |
|---|---|---|---|---|---|---|---|---|
| A | 2.0 | 2.616 | Initial: 120.94 | Initial: 122.74 | Initial: 114.93 | Initial: 17.117 | -336284.29 | -336290.34 |
|   | 2.0 | 2.619 | Final: 103.66 | Final: 131.41 | Final: 124.86 | Final: 17.117 | | |
| B | 2.0 | 2.547 | Initial: 120.94 | Initial: 122.74 | Initial: 114.93 | Initial: 17.117 | -336288.93 | -336290.37 |
|   | 2.0 | 2.510 | Final: 106.76 | Final: 137.36 | Final: 115.67 | Final: 17.117 | | |
| C | 2.0 | 2.556 | Initial: 120.94 | Initial: 122.74 | Initial: 114.93 | Initial: 17.117 | -336288.54 | -336290.31 |
|   | 2.0 | 2.506 | Final: 105.71 | Final: 131.86 | Final: 122.41 | Final: 17.117 | | |
| D | 2.0 | 2.526 | Initial: 120.94 | Initial: 122.74 | Initial: 114.93 | Initial: 17.117 | -336287.90 | -336290.31 |
|   | 2.0 | 2.534 | Final: 106.01 | Final: 128.56 | Final: 125.47 | Final: 17.117 | | |
| E | 2.475 | 2.503 | Initial: 120.94 | Initial: 122.74 | Initial: 114.93 | Initial: 17.592 | -336289.66 | -336290.29 |
|   | 2.0 | 2.476 | Final: 113.43 | Final: 124.82 | Final: 119.98 | Final: 17.592 | | |
| F | 2.0 | 2.611 | Initial: 114.68 | Initial: 123.50 | Initial: 120.030 | Initial: 16.687 | -336283.67 | -336290.31 |
|   | 2.0 | 2.610 | Final: 98.988 | Final: 131.80 | Final: 129.244 | Final: 16.687 | | |

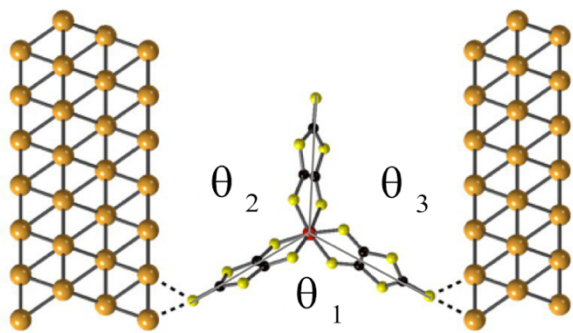

**Figure S1:** Depiction of the molecular junction illustrating the angles between the long axis of the dithiolate ligands.

2. Theoretical methodology

Transport calculations are performed using the SMEAGOL code[2] that interfaces the non-equilibrium Green's function (NEGF) approach to electron transport with the density functional theory (DFT) package SIESTA[3]. Each structural optimization calculation was performed using the original SIESTA code.

In our simulations the transport junction is constructed by placing the molecule between two Au(111)-oriented surfaces with 7x7 cross section. This mimics a standard transport break-junction experiment with the most used gold surface orientation. The choice of a gold electrode arises from the stability of the sulfur-gold bond that ensures the best attachment between the molecule and the surfaces.

Thus, the structures are relaxed until the maximum forces are less than 0.01 eV/Å. A real space grid with and equivalent plane wave cutoff of 200 Ry (enough to ensure convergence) has been used to calculate the various matrix elements. Finally, the electronic temperature of the calculation (unless specified) is set to 0.1 K to mimic low-temperature limit conditions. During the calculation, the total system is divided in three parts: a left-hand side lead, a central scattering region (SR) and a right-hand side lead. The scattering region contains the molecule as well as four atomic layers of each lead, which are necessary to relax the electrostatic potential to the bulk level of Au.

The convergence of the electronic structure of the leads is achieved with 2x2x128 Monkhorst-Pack k-point mesh, while for the SR one sets open boundary conditions in the transport direction and periodic ones along the transverse plane, for which an identical k-point mesh is used (2x2x1 k-points). The exchange-correlation potential is described by the Ceperley-Alder local density approximation (LDA)[4] as implemented for using the ASIC approach (see below). One of the main problems using local/semilocal DFT functionals for the study of the interfaces between inorganic leads and molecules is the self-interaction error. This leads to an over-delocalization of the electronic density producing the HOMO orbitals to be set higher in energy artificially. The LUMO orbitals could also be found at lower energy values that the commonly expected. In order to avoid this problem, we have used the Atomic Self Interaction Correction (ASIC) method[5].

This methodology is known to correctly set the position of the HOMO orbital, and it has also been shown to improve the energy level alignment when the junction is formed, leading to values of conductance in better agreement with experiments[6].

The Au-valence electrons are represented over a numerical s-only single-θ basis set that has been previously demonstrated to offer a good description of the energy region around the Fermi level[7]. In contrast, for the other atoms (S, C and V) we use a full-valence double-θ basis set with polarization (basis size was increased until convergence). Norm-conserving Troullier-Martins pseudopotentials[8] (previously adapted to ASIC methodology) are employed to describe the core-electrons in all the cases.

Finally, the spin-dependent current, $I_\sigma$, flowing through the junction is calculated from the Landauer-Büttiker formula[9],

$$I_\sigma(V) = \frac{e}{h} \int_{-\infty}^{+\infty} T_\sigma(E,V)[f_L(E - \mu_L) - f_R(E - \mu_R)]dE$$

Where the total current $I_{tot}$, is the sum of both the spin-polarized components, $I_\sigma$, where $\sigma$ = spin-up/spin-down. Here $T_\sigma(E,V)$ is the transmission coefficient[2], and $f_{L/R}$ are the Fermi functions associated with the two electrodes' chemical potentials, $\mu_{L/R} = \mu_\sigma \pm V/2$, where $\mu_\sigma$ is the electrodes' common Fermi level.

In our two-spin-fluid approximation (there is no spin-flip mechanism) majority and minority spins carry two separate spin currents, and the resultant current spin polarization, SP, is calculated as

$$SP = \frac{I_{up} - I_{down}}{I_{up} + I_{down}}$$

Note that although this approach is widely used to perform theoretical predictions within the spintronic community, [10] it has some limitations inherent to the basics of DFT methodology. One may note that theoretical spin asymmetry corresponds to what is expected from open-shell DFT calculation in a magnetic molecule. Thus, within the mono-determinant approximation of DFT only a concrete $M_S$ state is calculated. In literature this has been sometimes confused with the obtainment of a super-efficient spin filter. Contrary, in our case where we have a spin S=1/2 generated by the $V^{IV}$ atom it is possible to assume that such mono determinantal description is valid when the low temperature guarantees the spin orientation and it is experimentally fixed by an external polarizing magnetic field. Note that in this case, no explicit magnetic field is included in the calculation.